\documentclass[prd,twocolumn,amsmath,amssymb]{revtex4}
\usepackage{amssymb}
\usepackage{amsfonts}
\usepackage{graphicx}
\usepackage{keyval,graphicx}
\usepackage{textcomp,wasysym}

\begin{document}
\title{\boldmath Measurements of the observed cross sections for $e^+e^-
\to $ {\it  exclusive light hadrons}  containing $K_S^0$ meson at
$\sqrt s=3.773$ and 3.650 GeV }

\author{
\small M.~Ablikim$^{1}$,              J.~Z.~Bai$^{1}$, Y.~Bai$^{1}$,
Y.~Ban$^{11}$, X.~Cai$^{1}$, H.~F.~Chen$^{15}$, H.~S.~Chen$^{1}$,
H.~X.~Chen$^{1}$, J.~C.~Chen$^{1}$, Jin~Chen$^{1}$
% ...
\\
\vspace{0.2cm}
(BES Collaboration)\\
\vspace{0.2cm}
{\it
$^{1}$ Institute of High Energy Physics, Beijing 100049, People's Republic of China\\
% ...
$^{11}$ Peking University, Beijing 100871, People's Republic of China\\
% ...
$^{15}$ University of Science and Technology of China, Hefei 230026,
People's Republic of China\\
% ...
}
}

\begin{abstract}
By analyzing the data sets of 17.3 pb$^{-1}$ taken at $\sqrt s=
3.773$ GeV and of 6.5 pb$^{-1}$ taken at $\sqrt s= 3.650$ GeV with
the BES-II detector at the BEPC collider, we measure the observed
cross sections for the exclusive light hadron final states of
$K_S^0K^-\pi^+$, $K_S^0K^-\pi^+\pi^0$, $K_S^0K^-\pi^+\pi^+\pi^-$,
$K_S^0K^-\pi^+\pi^+\pi^-\pi^0$, $K_S^0K^-\pi^+\pi^+\pi^+\pi^-\pi^-$
and $K_S^0K^-\pi^+\pi^0\pi^0$ produced in $e^+ e^-$ annihilation at
the two energy points. We set the upper limits on the observed cross
sections and the branching fractions for $\psi(3770)$ decay to these
final states at $90\%$ C.L..
\end{abstract}

\maketitle

\oddsidemargin -0.2cm
\evensidemargin -0.2cm

\section{\bf INTRODUCTION}

The $\psi(3770)$ resonance is expected to decay almost entirely into
$D\bar D$ meson pairs since its width is almost two orders of
magnitude larger than that of $\psi(3686)$ \cite{prl39_526}. In
recent years, the study of the $\psi(3770)$ non-$D\bar D$ decays
becomes an attractive study field in the charmonium energy region
due to the existing puzzle that about $38\%$ of $\psi(3770)$ does
not decay into $D\bar D$ meson pairs \cite{hepex_0506051}. To
understand the possible excess of the $\psi(3770)$ cross section
relative to the $D\bar D$ cross section, BES and CLEO Collaborations
made many efforts to study the $\psi(3770)$ non-$D\bar D$ decays.
The CLEO Collaboration measured the $e^+e^-\to\psi(3770)\to$
non-$D\bar D$ cross section to be $(-0.01\pm0.08^{+0.41}_{-0.30})$
nb \cite{prl96_092002}. While the BES Collaboration measured the
branching fraction for $\psi(3770)\to$ non$-D\bar D$ decay to be
$(15\pm5)\%$
\cite{plb641_145,prl97_121801,plb659_74,prd76_000000,pdg07}, which
indicates that, contrary to what is generally expected, the
$\psi(3770)$ might substantially decay into non$-D \bar D$ final
states or there are some new structure or physics effects which may
partially be responsible for the largely measured non-$D\bar D$
branching fraction of the $\psi(3770)$ decays
\cite{prl101_102004,plb668_263}. BES Collaboration observed the
first non$-D \bar D$ decay mode for $\psi(3770) \to
J/\psi\pi^+\pi^-$, and measured its decay branching fraction to be
${\mathcal B}[\psi(3770) \to J/\psi\pi^+\pi^-]=
(0.34\pm0.14\pm0.09)\%$ \cite{hepnp28_325,plb605_63}. This was
confirmed by CLEO Collaboration \cite{prl96_082004}. Latter, CLEO
Collaboration observed more $\psi(3770)$ exclusive non-$D\bar D$
decays, $\psi(3770)\to J/\psi\pi^0\pi^0$, $J/\psi\pi^0$,
$J/\psi\eta$ \cite{prl96_082004}, $\gamma\chi_{cJ}(J=0,1,2)$
\cite{prl96_182002,prd74_031106} and $\phi\eta$ \cite{prd74_012005},
etc. Summing over these measured branching fractions yields the sum
of the branching fractions for the $\psi(3770)$ exclusive non-$D\bar
D$ decays not more than 2\%. In addition, BES and CLEO
Collaborations also attempted to search for other $\psi(3770)$
exclusive charmless decays
\cite{prd70_077101,prd72_072007,plb650_111,plb656_30,epjc52_805}
\cite{prd74_012005,prl96_032003,prd73_012002}. However, the existing
results can not clarify the possible excess. For better
understanding the origin of the possible excess, search for more
$\psi(3770)$ exclusive charmless decays will be helpful.

In this Letter, we report measurements of the observed cross
sections for the exclusive light hadron final states of
$K_S^0K^-\pi^+$ (Throughout the Letter, charge conjugation is
implied), $K_S^0K^-\pi^+\pi^0$, $K_S^0K^-\pi^+\pi^+\pi^-$,
$K_S^0K^-\pi^+\pi^+\pi^-\pi^0$, $K_S^0K^-\pi^+\pi^+\pi^+\pi^-\pi^-$
and $K_S^0K^-\pi^+\pi^0\pi^0$ at the center-of-mass energies of
3.773 and 3.650 GeV with the same method as the one used in our
previous works \cite{plb650_111,plb656_30,epjc52_805}. With the
measured cross sections at the two energy points, we set the upper
limits on the observed cross sections and the branching fractions
for $\psi(3770)$ decay to these final states. The measurements are
made by analyzing the data set of 17.3 pb$^{-1}$ collected at $\sqrt
s= 3.773$ GeV [called as the $\psi(3770)$ resonance data] and the
data set of 6.5 pb$^{-1}$ collected at $\sqrt s= 3.650$ GeV (called
as the continuum data) with the BESII detector at the BEPC collider.

\section{BESII detector}
The BES-II is a conventional cylindrical magnetic detector that is
described in detail in Refs. \cite{nima344_319,nima458_627}. A
12-layer vertex chamber (VC) surrounding the beryllium beam pipe
provides input to the event trigger, as well as coordinate
information. A forty-layer main drift chamber (MDC) located just
outside the VC yields precise measurements of charged particle
trajectories with a solid angle coverage of $85\%$ of 4$\pi$; it
also provides ionization energy loss ($dE/dx$) measurements which
are used for particle identification. Momentum resolution of
$1.7\%\sqrt{1+p^2}$ ($p$ in GeV/$c$) and $dE/dx$ resolution of
$8.5\%$ for Bhabha scattering electrons are obtained for the data
taken at $\sqrt s= 3.773$ GeV. An array of 48 scintillation counters
surrounding the MDC measures the time of flight (TOF) of charged
particles with a resolution of about 180 ps for electrons. Outside
the TOF, a 12 radiation length, lead-gas barrel shower counter
(BSC), operating in limited streamer mode, measures the energies of
electrons and photons over $80\%$ of the total solid angle with an
energy resolution of $\sigma_E/E=0.22/\sqrt{E}$ ($E$ in GeV) and
spatial resolutions of $\sigma_{\phi}=7.9$ mrad and $\sigma_z=2.3$
cm for electrons. A solenoidal magnet outside the BSC provides a 0.4
T magnetic field in the central tracking region of the detector.
Three double-layer muon counters instrument the magnet flux return
and serve to identify muons with momentum greater than 500 MeV/$c$.
They cover $68\%$ of the total solid angle.

\section{ EVENT SELECTION}
\label{evtsel}

In the reconstruction of the $K_S^0K^-\pi^+$, $K_S^0K^-\pi^+\pi^0$,
$K_S^0K^-\pi^+\pi^+\pi^-$, $K_S^0K^-\pi^+\pi^+\pi^-\pi^0$,
$K_S^0K^-\pi^+\pi^+\pi^+\pi^-\pi^-$ and $K_S^0K^-\pi^+\pi^0\pi^0$
final states, the $K^0_S$ and $\pi^0$ mesons are reconstructed
through the decays of $K^0_S\to \pi^+\pi^-$ and $\pi^0\to
\gamma\gamma$.

For each candidate event, we require that at least four charged
tracks are well reconstructed in the MDC with good helix fits, and
the polar angle of each charged track satisfies $|\rm
cos\theta|<0.85$. The charged tracks (except for the $K_S^0$ meson
reconstruction) are required to originate from the interaction
region $V_{xy}<2.0$ cm ($V_{xy}<8.0$ cm) and $|V_z|<20.0$ cm, where
$V_{xy}$ and $|V_z|$ are the closest approaches in the $xy$-plane
and the $z$ direction, respectively.

The charged particles are identified by using the $dE/dx$ and TOF
measurements, with which the combined confidence levels $CL_{\pi}$
and $CL_K $ for pion and kaon hypotheses are calculated. The pion
and kaon candidates are required to satisfy $CL_{\pi}>0.001$ and
$CL_K >CL_{\pi}$, respectively. To reconstruct $K^0_S$ mesons, we
require that the $\pi^+\pi^-$ meson pairs must originate from a
secondary vertex which is displaced from the event vertex at least
by 4 mm in the $xy$-plane.

The photons are selected with the BSC measurements. The good photon
candidates are required to satisfy the following criteria: the
energy deposited in the BSC is greater than 50 MeV, the
electromagnetic shower starts in the first 5 readout layers, the
angle between the photon and the nearest charged track is greater
than $22^{\circ}$ \cite{plb_597_39,plb_608_24}, and the opening angle between
the cluster development direction and the photon emission direction
is less than $37^{\circ}$ \cite{plb_597_39,plb_608_24}.

For each candidate event, there may be several different charged
and/or neutral track combinations satisfying the above selection
criteria for exclusive light hadron final states. Each combination
is subjected to an energy-momentum conservation kinematic fit.
For the processes
containing $\pi^0$ meson in the final states, an additional
constraint kinematic fit is imposed on $\pi^0\to\gamma\gamma$.
Candidates with a fit probability larger than 1$\%$ are accepted. If
more than one combination satisfies the selection criteria in an
event, only the combination with the longest decay distance of
the reconstructed $K_S^0$ mesons is retained.

To suppress the background from the $D\bar D$ decays, we use the
double tag method \cite{npb727_395} to remove the $D\bar D$ events.
For example, for the $K_S^0 K^-\pi^+\pi^0$ final state, we exclude
the all possible events from $D\bar D$ decays by rejecting those in
which the $D$ and $\bar D$ mesons can be reconstructed in the decay
modes of $D^-\to K_S^0K^-$ and $D^+\to \pi^+\pi^0$, $D^-\to
K^-\pi^0$ and $D^+\to K_S^0 \pi^+$, $\bar D^0\to K_S^0 \pi^0$ and
$D^0 \to K^-\pi^+$ \cite{npb727_395}. For the other final states,
the events from $D\bar D$ decays are suppressed similarly. The
remaining contaminations from $D\bar D$ decays due to particle
misidentification or missing photon(s) are accounted by using Monte
Carlo simulation, as discussed in Section \ref{backsub}.

\section{DATA ANALYSIS}

\begin{figure}[htbp]
\begin{center}
  \includegraphics[width=8cm,height=7cm]
{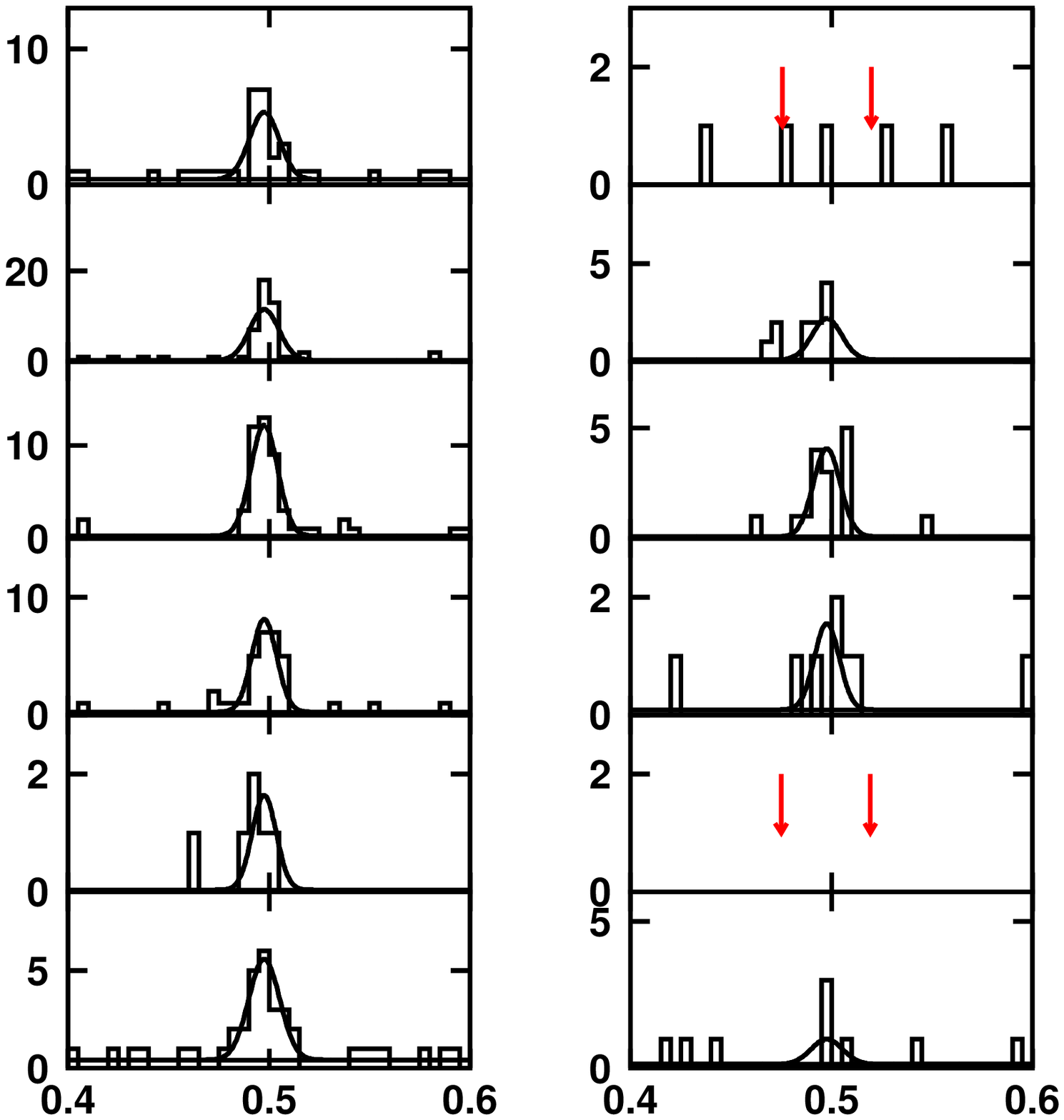}
      \put(-145,178){\small \bf (a)}
      \put(-50,178){\small \bf (a')}
      \put(-145,150){\small \bf (b)}
      \put(-50,150){\small \bf (b')}
      \put(-145,122){\small \bf (c)}
      \put(-50,122){\small \bf (c')}
      \put(-145,94){\small \bf (d)}
      \put(-50,94){\small \bf (d')}
      \put(-145,66){\small \bf (e)}
      \put(-50,66){\small \bf (e')}
      \put(-145,38){\small \bf (f)}
      \put(-50,38){\small \bf (f')}
      \put(-165,5){\bf Invariant mass (GeV/c$^2)$}
      \put(-220,60){\normalsize \rotatebox{90}{Events/0.005GeV/c$^2$}}
\caption{
The $\pi^+\pi^-$ invariant mass spectra of the candidates for the
(a) $K_S^0K^-\pi^+$,
(b) $K_S^0K^-\pi^+\pi^0$,
(c) $K_S^0K^-\pi^+\pi^+\pi^-$,
(d) $K_S^0K^-\pi^+\pi^+\pi^-\pi^0$,
(e) $K_S^0K^-\pi^+\pi^+\pi^+\pi^-\pi^-$ and
(f) $K_S^0K^-\pi^+\pi^0\pi^0$ final states
selected from the $\psi(3770)$ resonance data (left)
and the continuum data (right).}
\label{fig:data:con}
\end{center}
\end{figure}

In the data analysis, these processes containing $K^0_S$ meson in
the final state are studied by examining the invariant mass spectra
of the $\pi^+\pi^-$ combinations satisfying the above selection
criteria for the $K^0_S$ meson reconstruction. The invariant masses
of the $\pi^+\pi^-$ combinations are calculated with the momentum
vectors from the $K^0_S$ reconstruction. Figure \ref{fig:data:con}
shows the resulting distribution of the invariant masses of the
$\pi^+\pi^-$ combinations from the selected candidates for the
$K_S^0K^-\pi^+$, $K_S^0K^-\pi^+\pi^0$, $K_S^0K^-\pi^+\pi^+\pi^-$,
$K_S^0K^-\pi^+\pi^+\pi^-\pi^0$, $K_S^0K^-\pi^+\pi^+\pi^+\pi^-\pi^-$
and $K_S^0K^-\pi^+\pi^0\pi^0$ final states. In each figure, the peak
around the $K^0_S$ nominal mass indicates the production of
$e^+e^-\to$ exclusive light hadrons containing $K_S^0$ meson.
Fitting the $\pi^+\pi^-$ invariant mass spectra with a Gaussian
function for the $K^0_S$ signal and a flat background yields the
number of the events for each process observed from the $\psi(3770)$
resonance data and the continuum data. In the fit, the $K^0_S$ mass
and its mass resolution are fixed at the values obtained by
analyzing Monte Carlo samples.

\section{BACKGROUND SUBTRACTION}
\label{backsub}

Some other events may contribute to the selected candidate events
for $e^+e^- \to f$ ($f$ represents exclusive light hadron final
state). These include the events from $J/\psi$ and $\psi(3686)$
decays due to ISR returns, the events from the other final states
due to misidentifying a pion as a kaon or reverse, and the events
from $D\bar D$ decays. The number $N^{\rm b}$ of these
contaminations should be subtracted from the number $N^{\rm obs}$ of
the candidates for $e^+e^- \to f$. The estimation of them can be
done based on Monte Carlo simulation. The details about the
background subtraction have been described in Ref.
\cite{plb650_111}. For each background channel except $D\bar D$
decays, 50,000 or 100,000 Monte Carlo events are used in the
background estimation. The Monte Carlo sample of each different
background channel is from ten to several thousands times larger
than the data in size.

Monte Carlo study shows that the contaminations from $\psi(3770) \to
J/\psi\pi^+\pi^-$, $\psi(3770) \to J/\psi\pi^0\pi^0$, $\psi(3770)
\to J/\psi\pi^0$ and $\psi(3770) \to\gamma \chi_{cJ}\hspace{0.1cm}
(J=0,1,2)$ can be neglected.

Even though we have removed the main contaminations from $D\bar D$
decays in the previous event selection (see section \ref{evtsel}),
there are still some events from $D\bar D$ decays satisfying the
selection criteria for the light hadron final states due to particle
misidentification or missing photon(s). The number of these
contaminations from $D\bar D$ decays are further removed
by analyzing a Monte Carlo sample which is about forty times larger
than the $\psi(3770)$ resonance data. The Monte Carlo events are
generated as $e^+e^- \to D\bar D$ at $\sqrt s=$ 3.773 GeV,
where the $D$ and $\bar D$ mesons are set to decay into all
possible final states with the branching fractions quoted from PDG
\cite{pdg07}.

Subtracting the number $N^{\rm b}$ of these contaminations from the
number $N^{\rm obs}$ of the candidate events, we obtain the net
number $N^{\rm net}$ of the signal events for each process. For the
$K_S^0K^-\pi^+$ and $K_S^0K^-\pi^+\pi^+\pi^+\pi^-\pi^-$ final
states, for which only a few signal events are observed from the
continuum data, we set the upper limits $N^{\rm up}$ on the number
of the signal events at 90\% C.L.. Here, we use the Feldman-Cousins
method \cite{prd57_3873} and assume that the background is absent.
The numbers of $N^{\rm obs}$, $N^{\rm b}$ and $N^{\rm net}$ (or
$N^{\rm up}$) are summarized in the second, third and fourth columns
of Tabs. \ref{tab:data} and \ref{tab:con}.
For each process, the background events in the $\psi(3770)$ resonance data
are dominant by $D\bar D$ decays and $\psi(3686)$ decays. While, there is
no $D\bar D$ decay in the continuum data, and the $\psi(3686)$ production
cross section at $\sqrt s=$ 3.650 GeV is much less than that at $\sqrt s=$
3.773 GeV. So, the number of the background events in the continnum data
can almost be negligible.

\section{ RESULTS}
\label{results}

\subsection{ Monte Carlo efficiency}

To estimate the detection efficiency $\epsilon$ for $e^+e^-\to f$,
we use a phase space generator including initial state radiation and
vacuum polarization corrections \cite{yf41_377} with $1/s$ energy
dependence in cross section. Final state radiation \cite{cpc79_291}
decreases the detection efficiency not more than 0.5\%. Detailed
analysis based on Monte Carlo simulation for the BES-II detector
\cite{nima552_344} gives the detection efficiencies for each process
at $\sqrt s= 3.773$ and 3.650 GeV, which are summarized in the fifth
columns of Tabs. \ref{tab:data} and \ref{tab:con}, where the
detection efficiencies do not include the branching fractions for
$K_S^0 \to \pi^+\pi^-$ and $\pi^0\to\gamma\gamma$, ${\mathcal
B}(K_S^0 \to \pi^+ \pi^-)$ and ${\mathcal B}(\pi^0 \to
\gamma\gamma)$.

\subsection{Observed cross sections}

Let ${\mathcal B}_{\pi^0}$ = ${\mathcal B}(\pi^0 \to \gamma \gamma)$
for the modes of $K_S^0K^-\pi^+\pi^0$ and
$K_S^0K^-\pi^+\pi^+\pi^-\pi^0$, ${\mathcal B}_{\pi^0}$ = ${\mathcal
B}^2(\pi^0 \to \gamma\gamma)$ for the mode of
$K_S^0K^-\pi^+\pi^0\pi^0$ and ${\mathcal B}_{\pi^0}$ = 1 for
$K_S^0K^-\pi^+$, $K_S^0K^-\pi^+\pi^+\pi^-$ and
$K_S^0K^-\pi^+\pi^+\pi^+\pi^-\pi^-$, where ${\mathcal B}(\pi^0 \to
\gamma\gamma)$ is the branching fraction for the decay of $\pi^0 \to
\gamma\gamma$, then the observed cross section for $e^+e^- \to f$
can be determined by
\begin{equation}
\sigma_{e^+e^- \to f}=\frac{N^{\rm net}} {{\mathcal L} \times
\epsilon \times {\mathcal B}(K_S^0 \to \pi^+ \pi^-) \times {\mathcal
B}_{\pi^0}}, \label{eq:sig:net}
\end{equation}
where ${\mathcal L}$ is the integrated luminosity of the data set,
$N^{\rm net}$ is the number of the signal events, $\epsilon$ is the
detection efficiency and ${\mathcal B}(K_S^0 \to \pi^+ \pi^-)$ is
the branching fraction for the decay of $K_S^0 \to \pi^+ \pi^-$.
Inserting these numbers in Eq. (\ref{eq:sig:net}), we obtain the
observed cross sections for each process at $\sqrt s= 3.773$ and
3.650 GeV. They are summarized in Tabs. \ref{tab:data} and
\ref{tab:con}, where the first error is statistical and the second
systematic. In the measurements of the observed cross sections, the
systematic errors arise from the uncertainties in integrated
luminosity of the data set ($2.1\%$ \cite{plb641_145,prl97_121801}),
photon selection ($2.0\%$ per photon), tracking efficiency ($2.0\%$
per track), particle identification ($0.5\%$ per pion or kaon),
kinematic fit ($1.5\%$), $K_S^0$ reconstruction ($1.1\%$
\cite{plb_608_24}), branching fractions quoted from PDG
\cite{jpg33_1} ($0.03\%$ for ${\mathcal B}(\pi^0 \to \gamma\gamma)$
and $0.07\%$ for ${\mathcal B}(K_S^0 \to \pi^+ \pi^-)$), Monte Carlo
modeling ($6.0\%$ \cite{plb650_111,plb656_30,epjc52_805}), Monte
Carlo statistics ($1.4\%\sim4.4\%$), background subtraction
($0.0\%\sim3.0\%$) and fit to mass spectrum ($0.4\%\sim 8.5\%$).
Adding these uncertainties in quadrature yields the total systematic
error $\Delta_{\rm sys}$ for each mode at $\sqrt s= 3.773$ and 3.650
GeV.

The upper limit $\sigma_{e^+e^- \to f}^{\rm up}$ on the observed
cross sections for the $K_S^0K^-\pi^+$ and
$K_S^0K^-\pi^+\pi^+\pi^+\pi^-\pi^-$ final states at $\sqrt s= 3.650$
GeV are set with Eq. (\ref{eq:sig:net}) by substituting $N^{\rm
net}$ with $N^{\rm up}/(1-\Delta_{\rm sys})$, where $N^{\rm up}$ is
the upper limit on the number of the signal event, and $\Delta_{\rm
sys}$ is the systematic error in the cross section measurement.
Inserting the corresponding numbers in the equation, we obtain the
upper limits on the observed cross sections for $e^+e^-\to
K_S^0K^-\pi^+$ and $e^+e^-\to K_S^0K^-\pi^+\pi^+\pi^+\pi^-\pi^-$ at
$\sqrt s= 3.650$ GeV, which are also listed in Tab. \ref{tab:con}.

\subsection{Upper limits on the observed cross sections and the
branching fractions for $\psi(3770) \to f $}

If we ignore the possible interference effects between the continuum
and resonance amplitudes, and the difference of the vacuum
polarization corrections at $\sqrt s= 3.773$ and 3.650 GeV, we can
determine the observed cross section $\sigma _{\psi(3770) \to f}$
for $\psi(3770)\to f$ at $\sqrt s=3.773$ GeV by comparing the
observed cross sections $\sigma _{e^+e^- \to f}^{3.773
\hspace{0.05cm}\rm GeV}$ and $\sigma _{e^+e^- \to
f}^{3.650\hspace{0.05cm}\rm GeV}$ for $e^+e^- \to f$ measured at
$\sqrt s= 3.773$ and 3.650 GeV, respectively. It can be given by
\begin{equation}
\sigma _{\psi(3770) \to f}= \sigma _{e^+e^- \to f}^{3.773
\hspace{0.05cm}\rm GeV} - f_{\rm co}  \times \sigma _{e^+e^- \to
f}^{3.650\hspace{0.05cm}\rm GeV}, \label{eq:sig:3770}
\end{equation}
where $f_{\rm co}=3.650^2/3.773^2$ is the normalization factor to
consider the $1/s$ cross section dependence. The results are
summarized in the second column of Tab. \ref{tab:psipp:f}, where the
first error is the statistical, the second is the independent
systematic arising from the uncertainties in the Monte Carlo
statistics, in the fit to the mass spectrum and in the background
subtraction, and the third is the common systematic error arising
from the other uncertainties as discussed in the subsection B.

The upper limit on the observed cross section $\sigma^{\rm
up}_{\psi(3770) \to f}$ for $\psi(3770) \to f$ at $\sqrt s=3.773$
GeV is set by shifting the cross section by 1.64$\sigma$, where
$\sigma$ is the total error of the measured cross section. The
results on $\sigma^{\rm up}_{\psi(3770) \to f}$ are summarized in
the third column of Tab. \ref{tab:psipp:f}.

The upper limit on the branching fraction ${\mathcal B}^{\rm
up}_{\psi(3770)\to f}$ for $\psi(3770)\to f$ is set by dividing its
upper limit on the observed cross section $\sigma^{\rm
up}_{\psi(3770)\to f}$ by the observed cross section $\sigma^{\rm
obs}_{\psi(3770)}=(7.15\pm0.27\pm0.27)$ nb \cite{plb650_111} for the
$\psi(3770)$ production at $\sqrt s=3.773$ GeV and a factor
$(1-\Delta \sigma^{\rm obs}_{\psi(3770)})$, where $\Delta
\sigma^{\rm obs}_{\psi(3770)}$ is the relative error of the
$\sigma^{\rm obs}_{\psi(3770)}$. The results on ${\mathcal B}^{\rm
up}_{\psi(3770)\to f}$ are summarized in the last column of Tab.
\ref{tab:psipp:f}.

\section{SUMMARY}

In this Letter, we present the measurements of the observed cross
sections for $K_S^0K^-\pi^+$, $K_S^0K^-\pi^+\pi^0$,
$K_S^0K^-\pi^+\pi^+\pi^-$, $K_S^0K^-\pi^+\pi^+\pi^-\pi^0$,
$K_S^0K^-\pi^+\pi^+\pi^+\pi^-\pi^-$ and $K_S^0K^-\pi^+\pi^0\pi^0$
produced in $e^+ e^-$ annihilation at $\sqrt s= 3.773$ and 3.650
GeV. These cross sections are obtained by analyzing the data sets of
17.3 pb$^{-1}$ taken at $\sqrt s= 3.773$ GeV and of 6.5 pb$^{-1}$ at
$\sqrt s= 3.650$ GeV with the BES-II detector at the BEPC collider.
By comparing the observed cross sections for each process measured
at $\sqrt s= 3.773$ and 3.650 GeV, we set the upper limits on the
observed cross sections and the branching fractions for $\psi(3770)$
decay to these final states at $90\%$ C.L.. These measurements
provide helpful information to understand the mechanism of the
continuum light hadron production and the discrepancy between the
observed cross sections for $D\bar D$ and $\psi(3770)$ production.

\section{Acknowledgments}
The BES collaboration thanks the staff of BEPC for their hard
efforts. This work is supported in part by the National Natural
Science Foundation of China under contracts Nos. 10491300, 10225524,
10225525, 10425523, the Chinese Academy of Sciences under contract
No. KJ 95T-03, the 100 Talents Program of CAS under Contract Nos.
U-11, U-24, U-25, the Knowledge Innovation Project of CAS under
Contract Nos. U-602, U-34 (IHEP), the National Natural Science
Foundation of China under Contract  No. 10225522 (Tsinghua
University).

\begin{table*}[htbp]
\caption{The observed cross sections for $e^+e^-\to$ exclusive light
hadrons at $\sqrt s= 3.773$ GeV, where $N^{\rm obs}$ is the number
of events observed from the $\psi(3770)$ resonance data, $N^{\rm b}$
is the total number of background events, $N^{\rm net}$ is the
number of the signal events, $\epsilon$ is the detection efficiency,
$\Delta_{\rm sys}$ is the relative systematic error of the observed
cross section, and $\sigma$ is the observed cross section.}
\begin{center}
\small
\begin{tabular}{ l|l |l|l|l|c |l}  \hline
$e^+e^- \to$& $N_{\rm}^{\rm obs}$  & $N^{\rm b}$ &
$N^{\rm net}$ & $\epsilon$($\%$)& $\Delta_{\rm sys}$($\%$)  & $\sigma^{\rm obs}[{\rm pb}]$ \\
\hline
$K^0_S K^-\pi^+$                   & $18.4 \pm4.6$  & $0.1\pm 0.0$  & $18.3\pm4.6$   &$10.02\pm0.14$   &10.7    &$15.2\pm3.8\pm1.6$ \\
$ K^0_S K^-\pi^+\pi^0$             & $41.2 \pm6.6$  & $1.1\pm0.2$   & $40.1\pm6.6$   &$3.52\pm0.08$    &11.6    &$96.2\pm15.9\pm11.1$\\
$ K^0_S K^-\pi^+\pi^+\pi^-$        & $40.0 \pm 6.5$ &$1.0\pm0.2$    & $38.9\pm6.5$   &$3.56\pm0.06$    &14.2    &$91.5\pm15.3\pm13.0$ \\
$K^0_S K^-\pi^+\pi^+\pi^-\pi^0 $   & $24.5 \pm 5.2$ &$1.5\pm0.3$    & $23.0\pm5.2$   &$0.77\pm0.03$    &15.2    &$253.0\pm57.1\pm38.4$ \\
$K^0_SK^-\pi^+\pi^+\pi^+\pi^-\pi^-$& $4.8 \pm 2.2$  &$0.3\pm0.1$    & $4.5\pm2.2$    &$0.84\pm0.03 $   &18.4    &$44.4\pm21.9\pm8.2$ \\
$ K^0_S K^-\pi^+\pi^0\pi^0$        & $19.8 \pm4.9$  & $2.8\pm0.5$   & $17.0\pm4.9$   &$0.99\pm0.04$    &14.3    &$147.0\pm42.4\pm21.0$\\
\hline
\end{tabular}
\label{tab:data}
\end{center}
\end{table*}

\begin{table*}[htbp]
\caption{
The observed cross sections for $e^+e^-\to$ exclusive light hadrons at
$\sqrt s= 3.650$ GeV, where
$N^{\rm obs}$ is the number of events observed from the continuum data,
$N^{\rm up}$ is the upper limit on the number of the signal events,
$\sigma^{\rm up}$ is the upper limit on the observed cross section set at
90\% C.L.,
and the definitions of the other symbols are the same as those in Tab.
\ref{tab:data}.
}
\begin{center}
\small
\begin{tabular}{ l|c |l|c|l|c|l }  \hline
$e^+e^- \to$& $N_{\rm }^{\rm obs}$& $N^{\rm b}$ & $N^{\rm net}({\rm
or}\hspace{0.5em} N^{\rm up}_{\rm})$ & $\epsilon$($\%$)&
$\Delta_{\rm sys}$($\%$) & $\sigma^{\rm obs}(\sigma^{\rm up})[\rm pb]$ \\
\hline
$K^0_S K^-\pi^+$                    & $2$           &0.0   & $<5.91$        &$10.55\pm0.15$    &12.9    &$<14.3$  \\
$K^0_S K^-\pi^+\pi^0$               & $7.7\pm2.9$   &0.0   & $7.7 \pm2.9$   &$3.62\pm 0.09$    &11.6    &$47.9\pm18.0\pm5.6 $  \\
$K^0_S K^-\pi^+\pi^+\pi^-$          & $13.4 \pm3.8$ &0.0   & $13.4 \pm 3.8$ &$3.66\pm 0.06$    &14.1    &$81.4\pm23.1\pm11.5 $   \\
$K^0_S K^-\pi^+\pi^+\pi^-\pi^0 $    & $4.6\pm2.5$   &0.0   & $4.6\pm2.5$    &$0.87\pm 0.03$    &17.2    &$119.0\pm64.7\pm20.5$  \\
$K^0_SK^-\pi^+\pi^+\pi^+\pi^-\pi^-$ & $0$           &0.0   & $<2.44$        &$0.95\pm 0.03$    &18.1    &$<69.7$   \\
$K^0_S K^-\pi^+\pi^0\pi^0$          & $3.3\pm2.0$   &0.0   & $3.3\pm2.0$    &$1.12\pm 0.05$    &14.1    &$67.1\pm40.7\pm9.5$\\
\hline
\end{tabular}
\label{tab:con}
\end{center}
\end{table*}

\begin{table*}
\caption{The upper limits on the observed cross section $\sigma^{\rm
up}_{\psi(3770)\to f}$ at $\sqrt s=3.773$ GeV and the branching
fraction ${\mathcal B}^{\rm up}_{\psi(3770)\to f}$ for
$\psi(3770)\to f$ are set at 90\% C.L.. The $\sigma_{\psi(3770)\to
f}$ is calculated with Eq. (\ref{eq:sig:3770}), where the first
error is the statistical, the second is the independent systematic,
and the third is the common systematic error. The upper $^*$ denotes
that we neglect the contributions from the continuum production.}
\begin{center}
\small
\begin{tabular}{ l|l |c|l }  \hline
Decay Mode  &$\sigma_{\psi(3770) \to f}$ (pb)
            &$\sigma^{\rm up}_{\psi(3770) \to f}$(pb)
            & ${\mathcal B}^{\rm up}_{\psi(3770) \to f}$ \\ \hline
$K_S^0K^-\pi^+$                     &$15.2\pm3.8\pm0.2\pm1.6^*$     &$<22.0$    &$3.2\times 10^{-3}$\\
$K_S^0K^-\pi^+\pi^0$                &$51.4\pm23.2\pm2.6\pm5.8$      &$<90.7$    &$13.3\times 10^{-3}$\\
$K_S^0K^-\pi^+\pi^+\pi^-$           &$15.3\pm26.5\pm2.2\pm2.1$      &$<59.0 $   &$8.7  \times 10^{-3}$\\
$K_S^0K^-\pi^+\pi^+\pi^-\pi^0$      &$141.6\pm83.2\pm14.7\pm20.7$   &$<284.3$   &$41.8  \times 10^{-3}$\\
$K_S^0K^-\pi^+\pi^+\pi^+\pi^-\pi^-$ &$44.4\pm21.9\pm2.1\pm7.9^*$    &$<82.7$    &$12.2  \times 10^{-3}$\\
$K_S^0K^-\pi^+\pi^0\pi^0$           &$84.2\pm57.2\pm8.3\pm11.2$     &$<180.4$   &$26.5  \times 10^{-3}$\\
\hline
\end{tabular}
\label{tab:psipp:f}
\end{center}
\end{table*}

\end{document}